\documentclass[epj,nopacs]{svjour}
\usepackage{graphicx}
\usepackage{color}
\usepackage{multirow}
\usepackage{amsmath}
\usepackage{appendix}
\usepackage[colorlinks,linkcolor=red,anchorcolor=blue,citecolor=blue]{hyperref}
\usepackage{cite}
\begin{document}
\title{Probing Supermassive Black Hole Binaries with Orbital Resonances of Laser-Ranged Satellites}
% \subtitle{Do you have a subtitle?\\ If so, write it here}
\author{Minghui Du\inst{1} \and Qiong Deng\inst{2,1}\thanks{\emph{Co-first author.}} \and Yifan Bian\inst{2} \and Ziren Luo \inst{1,3,4} \and Peng Xu\inst{1,2,3,4}\thanks{\emph{Corresponding author. E-mail:} xupeng@imech.ac.cn}%
}                     % Do not remove
%
% \offprints{}          % Insert a name or remove this line
%
\institute{Center for Gravitational Wave Experiment, National Microgravity Laboratory, Institute of Mechanics, \\ Chinese Academy of Sciences, Beĳing 100190, China. 
\and Lanzhou Center of Theoretical Physics, Lanzhou University, Lanzhou 730000, China.
\and Taiji Laboratory for Gravitational Wave Universe (Beijing/Hangzhou), \\ University of Chinese Academy of Sciences, Beijing 100049, China.
\and Hangzhou Institute for Advanced Study, University of Chinese Academy of Sciences, Hangzhou 310124, China.}
\date{Received: date / Revised version: date}
\abstract{
  Coalescing supermassive black hole binaries (SMBHBs) are the primary source candidates for low frequency gravitational wave (GW) detections, which could bring us deep insights into galaxy evolutions over cosmic time and violent processes of spacetime dynamics. Promising candidates had been found based on optical and X-ray observations, which claims for new and ready-to-use GW detection approaches before the operations of space-borne antennas. We show that, satellite laser ranging (SLR) missions could serve as probes of coalescing SMBHBs through the GW-induced resonant effects. Lasting and characteristic imprints caused by such resonances in the residual distances or accelerations from SLR measurements are studied, and the detection SNR is analyzed with both the current and future improved ranging precisions. Within redshift $z \sim 1$, the threshold SNR=5 requires 1-2 years of accumulated data for the current precision and months of data for improved precision, which are workable for the data processing of SLR missions. Meanwhile, joint detections with multiple SLR missions could further improve the total SNR and the confidence level.  Such a detection scheme could fulfill the requirement of a tentative SMBHB probe during the preparing stage of LISA and Taiji, and it requires no further investment to any new and advanced facilities. It is also worthwhile to look back and re-process the archived data from the past decades, in where resonant signals from SMBHBs might be hidden. 
\PACS{
      {PACS-key}{discribing text of that key}   \and
      {PACS-key}{discribing text of that key}
     } % end of PACS codes
} %end of abstract
\maketitle

\section{Introduction}
\label{intro}
Ever since the landmark event GW150914 observed by Adv-LIGO~\cite{PhysRevLett.116.131103,LIGOScientific:2014pky}, the new window to the Universe had been opened by GW detections. 
With the follow-on observations of nearly one  hundred of events by the LIGO-VIRGO collaboration \cite{LIGOScientific:2020ibl,2021arXiv211103606T}, the new era of GW astronomy has gradually started off. 
To enclose the exciting sources of much larger and heavier astrophysical systems, one needs to explore the low frequency end of the GW spectrum with detectors of much longer baselines. 
In the next decade, the first generation space-borne antennas, including LISA (Laser Interferometer Space Antenna) ~\cite{2017arXiv170200786A} and the LISA-like Taiji mission~\cite{Hu2017TheTP,10.1093/ptep/ptaa083,LUO2020102918,2018arXiv180709495R} would be launched and cover the mHz band~\cite{Ruan:2020smc}. 
Mission concepts for decihertz band now include the DECIGO \cite{Kawamura:2020pcg}, AMIGO \cite{Ni:2021eqz} and TianQin ~\cite{TianQin:2015yph} projects. 
For the $\mu$Hz band, the more advanced space missions with baselines of Solar system size like the $\mu$Ares \cite{Sesana:2019vho} and ASTROD \cite{Ni:2012eh} were suggested.
Pulsar Timing Arrays (PTA) are reported to be promising probes for nHz GWs \cite{Lasky:2015lej,NANOGrav:2020bcs,Janssen:2014dka}. 
%Besides, Ref.~\cite{Fedderke:2021kuy,Fedderke:2020yfy,2022arXiv220407677F} proposed a series of mission concepts of using white dwarfs and inner Solar System asteroids as test masses to detects GWs around the $\mu$Hz and sub-$\mu$Hz bands. 
%While, these approaches have left the $10^{-5} - 10^{-4}$ Hz range unexplored in the foreseeable future.

Among the candidate sources within the low frequency range, coalescing SMBHBs ($10^6-10^9\ M_{\bigodot}$) are of the most exciting ones \cite{PhysRevD.93.024003}. 
Their collisions and mergers give rise to the most violent events in the visible Universe and produce the loudest GW signals within their frequency band. 
The knowledge of the population of coalescing SMBHBs and the detailed measurements of the entire wavetrains (inspiral, merger, and ringdown) will bring us deep insights into the growths and co-evolutions of the SMBHBs and their host galaxies, the expansion history of our Universe, the most violent dynamic behavior of curved spacetime and the nature of gravitation \cite{PhysRevD.93.024003,Will:2014kxa,Clifton:2011jh}. 
Therefore, coalescing SMBHBs are the primary candidates for many aforementioned missions. 
According to the roadmap of GW physics and astronomy \cite{bailes_gravitational-wave_2021}, LISA will firstly reveal the detailed information about SMBHBs till the 2030s.
On the other hand, SMBHB candidates could be identified by means of optical and X-ray observations. 
{\color{black}For example, Ref.~\cite{2022arXiv220111633J} claimed that the rapid decaying binary system SDSSJ1430+2303 discovered via optical and X-ray observations is expected to merge within 3 years, }which might provide us an excellent opportunity for multi-messenger observations if any low frequency GW detector was ready~\cite{Fan:2014kka,Fan:2017rkg}. 
Having the plentiful scientific objectives and the potential opportunities, it would be of great significance if any tentative detection of coalescing SMBHB could be made within this decade before the operations of LISA and Taiji missions.  

Similar to the case for a bound system of charged particles, which could have resonant interactions with incident electromagnetic waves when the wave frequency matches with the energy difference between certain states of the system, the response of a self-gravitating binary system to GW is especially evident when the frequency of GW matches a harmonic of the binary's orbital frequency, thereby inducing a resonant effect. 
Such resonant evolution of the orbit can accumulate in time, and may eventually enter the scope of possible detections. 
The studies of resonant responses of self-gravitating binary systems to incident GWs can be traced back to the 1970s \cite{1973ApL....14...51B,1975SvA....19..270R,1978ApJ...223..285M,1979ApJ...233..685T,1981ApJ...246..569M}, and had raised more concerns in these years ~\cite{Chicone:1995nn,Chicone:1996yc,Chicone:1999sg,PhysRevD.87.084009,2014SOR..2014E...1I,PhysRevLett.118.261102,Desjacques:2020fdi,Armaleo:2020yml,Blas:2019hxz,Iorio:2021cxt}. 
Recently, Blas and Jenkins had made important progress for this detection scheme and developed powerful tools to estimate the resonant evolutions of the binary orbital elements under stochastic GW backgrounds \cite{PhysRevD.105.064021}. 
Sensitivities to such resonant effects from stochastic backgrounds in SLR, Lunar laser ranging (LLR) and PTA missions turn out to be promising \cite{Blas:2021mqw}, which approves the detectability of such scheme. 
%for bridging the $\mu$Hz frequency gap in the GW landscape by means of 

Most of the GW signals studied in the aforementioned literature are stochastic in nature, and still the resonant detection scheme for general deterministic and individual signals is not yet fully investigated. 
In this work, we aim to study the feasibility and prospect of detecting, in the circumstance where orbital resonance takes place, the deterministic GW signals from coalescing SMBHBs via the SLR technique of the present day and the next decade. 
The Earth-satellite distances can be measured precisely and continuously for SLR missions, which hence provides the long-term and faithful tracking of the orbital evolutions of the satellites \cite{Luceri2019,URSCHL20071515}. 
In fact, data from SLR missions, such as LAGEOS 1, 2 and LARES, had already been proved to be very useful in relativistic experiments~\cite{1986PhRvL..56..278C,2004Natur.431..958C,2007Natur.449...41C,2009SSRv..148...71C,Ciufolini:2012rx,2016CQGra..33d5004I,2019NatSR...915881C,Ciufolini:2016ntr,Ciufolini:2019ezb}. 
For GW detections, the orbital harmonics of all the in-orbit laser-range satellites~\cite{ILRS} could form a “comb” in the sub-mHz range, which makes it possible to capture the strong chirping signals from coalescing SMBHBs consecutively by most of the missions in operation.   
Based on previous studies~\cite{1961mcm..book.....B,doi:10.1119/1.10237,murray_dermott_2000,1979ApJ...233..685T,smart1960celestial}, and especially \cite{PhysRevD.105.064021}, we 
% adjust the equations of orbital perturbations to describe the resonant evolutions of osculate orbital elements induced by individual gravitational wavetrains. 
give analytical and numerical analysis of the resonant evolutions of osculate orbital elements induced by GWs from coalescing SMBHBs in SLR missions and their dependence on relevant parameters.
An important finding is that, when such resonance takes place, a characteristic signature is left in the orbital evolutions of the laser-ranged satellites.
With the precision and multi-year data of orbit tracking, the resonance signal could be recovered with sophisticated data analysis methods.
Given the joint detections of each individual signal by different SLR missions, this would finally produce the high-confidence detection of a coalescing SMBHB. 

Limited by the precision of SLR measurements in present-days and in the near future, such detections may not give rise to detailed physical parameters of the sources, but the multi-year observations could probably provide us the first estimation of the population or event rate of coalescing SMBHBs within redshift $z\sim 0.1$.    
Such a detection scheme could fulfill the requirement of a tentative SMBHB probe in the decade before the launch of LISA (and Taiji), and it requires no further investment to any new and advanced facilities. 
The only efforts demanded will be the thorough analysis of the data, especially the re-analysis of the archived data from the past decades in where resonant signatures from coalescing SMBHBs might be hidden.

\section{Theoretical Tools}
For clarity, we will refer to the SMBHB as the ``source binary'' and the Earth-satellite system as the ``test binary''.
The phenomenon of orbital resonance can be described by the equations of motion (EoM) of the osculating orbital elements of the test binary, with GWs acting as small perturbations. 

We introduce a cylindrical coordinate $\{\boldsymbol{\hat{r}}, \boldsymbol{\hat{\theta}}, \boldsymbol{\hat{\ell}}\}$ whose origin is placed at the test binary's center of mass, and $\{\boldsymbol{\hat{r}}, \boldsymbol{\hat{\theta}}\}$ represent the bases of polar coordinates within the orbital plane, $\boldsymbol{\hat{\ell}}$ being the unit vector perpendicular to them. 
The unperturbed Keplerian motion of the satellite is characterized by six orbital elements $\boldsymbol{X} = \{P, e, I, \Omega, \omega, \varepsilon\}$, including the orbital period, eccentricity, inclination, longitude of ascending node, argument of pericenter, and the compensated mean anomaly.% see Fig. \ref{fig:orbit}.
% \begin{figure}
% \centering
% \includegraphics[width=.30\textwidth]{figures/orbit}
% \caption{The coordinates system for the Earth-satellite system, the orbit elements and the relative orientation of the source binary. \textit{\color{blue}maybe move this figure to appendix}}
% \label{fig:orbit}
% \end{figure}
% A schematic diagram illustrating the reference frame and orbital elements can be found in Fig. 1 of~\cite{PhysRevD.105.064021}.

In the absence of perturbations, the separation $r(t; \boldsymbol{X})$ of test binary, that is related to the most important observable of SLR missions, follows the Kepler's equations 
\begin{eqnarray}
    r(t; \boldsymbol{X}) &=& a\left[1 - e\cos E(t; \boldsymbol{X})\right], \label{eq:Unperturbedr}\\
    E(t; \boldsymbol{X}) &-& e\sin E(t; \boldsymbol{X}) = \frac{2\pi t}{P} + \varepsilon, \label{eq:UnperturbedE} \\
    \psi(t; \boldsymbol{X}) &=& 2\arctan \left[ \sqrt{\frac{1+e}{1-e}} \tan \frac{E(t; \boldsymbol{X})}{2} \right], \label{eq:Unperturbedpsi} 
\end{eqnarray}
where the true anomaly $\psi$ is defined as the angular position of the satellite measured from the pericenter, and $E$ is the so-called eccentric anomaly. 

With perturbations induced by incident GWs, Eq.(\ref{eq:Unperturbedr}-\ref{eq:Unperturbedpsi}) are valid under the condition that $\boldsymbol{X}$ are regarded as the osculating orbital elements, which varies with time satisfying the RTN-type Gaussian perturbation equations~\cite{SolarSystemAstrophysics,doi:10.1119/1.10237,murray_dermott_2000}. The effects of GWs, treated as small perturbations, can be written down in the form of Newtonian forces~\cite{1979ApJ...233..685T,10.1093/acprof:oso/9780198570745.001.0001}:
\begin{equation}
    \boldsymbol{F}_{\rm GW} = r(R\boldsymbol{\hat{r}} + T\boldsymbol{\hat{\theta}} + N\boldsymbol{\hat{\ell}}),
\end{equation}
where $R$, $T$ and $N$ are the perturbing forces per unit mass in the radial, tangential and normal directions relative to the orbit
\begin{eqnarray}
    R &=& \frac{1}{2}\ddot{h}_{ij}\hat{r}^i\hat{r}^j, \quad  T = \frac{1}{2}\ddot{h}_{ij}\hat{r}^i\hat{\theta}^j, \nonumber \\
    N &=& \frac{1}{2}\ddot{h}_{ij}\hat{r}^i\hat{\ell}^j, \quad h_{ij} = h_A e_{ij}^A.
\end{eqnarray}
$e_{ij}^A$ being the polarization tensors of the GW $ (A = +, \times)$. 
% In this paper we apply Einstein's sum rule to $A$, i.e. wherever $A$ is repeated, it should be summed over.
Finally, in terms of transfer function, the EoM of $\boldsymbol{X}$ can be written in a compact form~\cite{PhysRevD.105.064021}:
\begin{eqnarray}\label{eq:PerturbedOrbitalElements}
    \dot{\boldsymbol{X}} = \boldsymbol{T}^A(\boldsymbol{X}, \psi, \boldsymbol{\hat{n}}_{\rm GW}) \ddot{h}_A(\boldsymbol{\hat{n}}_{\rm GW}, t),
\end{eqnarray}
% Note that we have suppressed the dependence on GW source parameters for brevity. 
% The GW-induced perturbations depend on the orbital elements $\boldsymbol{X}$, and $\psi, \boldsymbol{\hat{n}}_{\rm GW}, t$, with $\boldsymbol{\hat{n}}_{\rm GW} = \boldsymbol{\hat{n}}_{\rm GW}(\vartheta, \phi)$ denoting the direction of source.} 
where $\boldsymbol{\hat{n}}_{\rm GW} = \boldsymbol{\hat{n}}_{\rm GW}(\vartheta, \phi)$ denotes the direction of source.
The transfer functions $\boldsymbol{T}^A$ define the linear responses of the orbital elements $\boldsymbol{X}$ to the incident GWs. Such linear responses are valid under the conditions that perturbations from GWs are small compared to that from Newtonian gravity and the back reactions from resonance to the incident GW field are ignorable. 
% The transfer functions $\boldsymbol{T}^A$ for the 6 orbital elements are
% \begin{eqnarray}\label{eq:TransferFuncP}
% 	T_P^A &=& \frac{3P^2 \gamma}{4\pi}\left(\frac{e\sin \psi}{1 + e \cos \psi} \hat{r}^i + \hat{\theta}^i\right) \hat{r}^j e^A_{ij}, \\
% 	T_e^A &=& \frac{\gamma^2 T_P^A}{3Pe} - \frac{P\gamma^5}{4\pi e}\frac{e^A_{ij}\hat{r}^i\hat{\theta}^j}{(1+e \cos\psi)^2}, \\
% 	T_I^A &=& \frac{P\gamma^3}{4\pi}\frac{\cos\theta}{(1 + e\cos\psi)^2}e^A_{ij}\hat{r}^i\hat{\ell}^j, \\
% 	T_\Omega^A &=& \frac{P\gamma^3}{4\pi \sin I}\frac{\sin \theta}{(1 + e\cos \psi)^2 e_{ij}^A\hat{r}^i\hat{\ell}^j}, \\
% 	T_\omega^A &=& \frac{P\gamma^3}{4\pi e} \left[\frac{\sin \psi (2 + e\cos \psi)}{(1 + e\cos \psi)^2} e_{ij}^A\hat{r}^i \hat{\theta}^j \right. \nonumber \\ & & - \left. \frac{\cos \psi e_{ij}\hat{r}^i \hat{r}^j}{1 + e\cos \psi}\right] - T_\Omega^A\cos I, \\
% 	T_\varepsilon^A &=& -\frac{P\gamma^4}{2\pi} \frac{e_{ij}^A \hat{r}^i \hat{r}^j}{(1 + e\cos\psi)^2} - \gamma \cos I T_\Omega^A - \gamma T_\omega^A.
% \end{eqnarray}
% {\color{black} where $\gamma \equiv \sqrt{1 - e^2}$, $\theta \equiv \psi + \omega$. $e^A_{ij}$ are the polarization tensors of GW, and $\hat{r}^i, \hat{\theta}^i, \hat{\ell}^i$ are the unit vectors of $\rm{r}, \rm{\theta}, \rm{\ell}$, respectively. 
The explicit forms of the transfer functions $\boldsymbol{T}^A$ and the coefficients of $e_{ij}^A$ in the test binary frame can be found in the work \cite{PhysRevD.105.064021}.
For SLR measurements, we are most concerned about the orbital period $P$ (or the semi-major axis $a$), which is directly related to the total energy of the test binary. The transfer function of $P$ can be expressed as
% For example, the most concerned element in this work i.e. the orbital period $P$ (or the semi-major axis $a$), is directly related to the total energy of the test binary, can be written down as 
\begin{eqnarray}\label{eq:TransferFuncP}
	T_P^A &=& \frac{3P^2 \gamma}{4\pi}\left(\frac{e\sin \psi}{1 + e \cos \psi} \hat{r}^i + \hat{\theta}^i\right) \hat{r}^j e^A_{ij},
\end{eqnarray}
where $\gamma \equiv \sqrt{1 - e^2}$.
% Note that in Eq.(\ref{eq:PerturbedOrbitalElements}) the post Newtonian correction to the orbital motion is neglected, since its irrelevant to resonance. 
% {\color{black} According to our estimate for the magnitude of resonance, the variation in the Earth-satellite system's total energy is subdominant compared to the energy of incident GW, thus the back-action of test binary to GW can be safely neglected.}
% In the following we will introduce the methods of solving Eq.(\ref{eq:PerturbedOrbitalElements}) numerically, and an analytical solution is acquired under several simplifications.

% Eq.(\ref{eq:PerturbedOrbitalElements}) is not closed until we include the time derivative of $\psi$. Following Refs.~\cite{smart1960celestial,1961mcm..book.....B} we have
% \begin{eqnarray}\label{eq:KeplerEquation}
%     \dot{\psi} %&=& \frac{\ell}{r^2} - \dot{\omega} - \dot{\Omega} \cos I \nonumber \\ 
%     &=& \frac{2\pi}{P} \frac{(1 + e\cos\psi)^2}{\gamma^3} - \dot{\omega} - \dot{\Omega} \cos I.
% \end{eqnarray}
{\color{black} Eqs.(\ref{eq:PerturbedOrbitalElements}) constitute a system of ODEs, which are solved numerically in the following analysis using the initial conditions $\boldsymbol{X}|_{t=0}=\boldsymbol{X}_0$. The orbital perturbation theory suggests that $\psi(t)$ can be calculated by solving the Kepler equations Eq.(\ref{eq:UnperturbedE}) and Eq.(\ref{eq:Unperturbedpsi}) with the orbital elements regarded as the osculating ones. }
% Without loss of generality, we further set $\psi = 0$ at $t = 0$, meaning that the binary is initially at its pericenter. As a result, the initial conditions (IC) are $\{\psi, \boldsymbol{X}\} = \{0, \boldsymbol{X}_0\}$ at $t = 0$.
% $$\left\{\begin{aligned}\label{eq:IC}
% t &= 0 \\
% \psi &= 0 \\
% \boldsymbol{X} &= \boldsymbol{X}_0
% \end{aligned}\right.$$

%\subsection{Analytical Solution}\label{sec:AnalyticalSolution}
To understand the resonant behavior qualitatively, we also derive an analytical solution under the simplifications that the orbit of test binary is nearly circular ($e \ll 1$) and the incident GW is modeled as a monochromatic wave with redshifted frequency $f_{\rm GW}$, initial phase $\varphi_{\rm GW}$ and amplitudes $H_A$. At the ``main'' resonance frequency $f_{\rm GW} = f_{\rm res} = 2 / P$, the variation of $P$ is dominated by a linear drift term. The secular perturbation of $P$, defined as $\dot{P}$ averaged over one orbit revolution, reads 
\begin{eqnarray}
    \dot{P}_{\rm sec} &=& 6\pi \gamma H_A\sqrt{G_{1A}^2 + G_{2A}^2} \nonumber \\ 
    & &\times \sin\left(\varphi_{\rm GW} - 2\varepsilon - \arctan \frac{G_{1A}}{G_{2A}} - \delta_{A\times}\frac{\pi}{2}\right),
\end{eqnarray}
where $G_{1A}$, $G_{2A}$ are constants determined by the angles of the GW source and the test binary (see Appendix \ref{sec:appendixA}), and $\delta_{A \times} = 1$ if $A = \times$ or $0$ if $A = +$.
In the case where the source binary and test binary are face-on, $\dot{P}_{\rm sec}$ can be expressed more concisely:
\begin{equation}
    \dot{P}_{\rm sec} = 12\pi \gamma H \sin(\varphi_{\rm GW} - 2\omega - 2\varepsilon),
\end{equation}
with $H \equiv H_A (\iota = 0)$.
Depending on the values of $\varphi_{\rm GW}$, $\omega$ and $\varepsilon$, $\dot{P}_{\rm sec}$ can be either positive or negative. 
Besides, two ``secondary'' resonances of order $\mathcal{O}(e)$ occur at $f_{\rm GW} = 1/P$ and $3/P$. The detailed derivation of this solution can be found in \ref{sec:appendixA}.
These discussions provide an intuitive demonstration of orbital resonance.
% and in analyzing the influence from the Milky Way DWD foreground they can be used to make an order-of-magnitude estimation. 

\section{{\color{black} An Example of merging SMBHB}}
{\color{black}We assume a SMBHB with the parameters of SDSSJ1430+2303 ~\cite{2022arXiv220111633J} as an example for the SMBHBs that will merge in the near future. Although the interpretation and detectability of SDSSJ1430+2303 are still under discussion, we are interested in a group of similar SMBHB systems instead of this specific one. This example will be referred to as our representative Target Source in the rest of this paper ({\bf TS} for short), and the variation of its parameters over a wide range will be discussed in~\ref{sec:AppendixB}.}
Among all the SLR missions, we take the laser ranging mission LAGEOS 2 ({\bf L2} for short) as a typical representative. The resonant responses of other SLR missions, including LAGEOS 1, LARES 1/2, Ajisai and ETALON 1/2 to the same GW signal are also discussed. The parameters of these missions can be found in~\cite{Pearlman,Ciufolini:2016ntr}. 
% since it has relatively large eccentricity (0.0135) compared to LAGEOS (0.0045) and LARES (0.0008), thus can be more representative for general cases with slightly eccentric orbits. 

The parameters relevant to our example are listed in Tab.~\ref{tab:Parameters}, where the initial values of $P, e, I$ for \textbf{L2} are taken from Ref.~\cite{Ciufolini:2016ntr}. 
Ref.~\cite{2022arXiv220111633J} reported the properties of the {\bf TS} constrained from electromagnetic observations.  It is proposed that this system is an uneven mass-ratio, highly eccentric SMBHB. 
While, at the frequencies of our interest, its orbit would be sufficiently circularized. 
Moreover, the components masses of the {\bf TS} are only determined with large uncertainty. 
Therefore, here we simply assume that it consists of two black holes with equal source-frame masses $M_{\rm bh} = 4\times 10^7 M_{\bigodot}$, and redshift $z = 0.08105$. 
The sky position and inclination of the {\bf TS}, as well as $\Omega$ and $\omega$ are randomly selected, since $\Omega$ and $\omega$ could change in time, and we are interested in a family of SMBHBs with properties similar to SDSSJ1430+2303 rather than this specific one. 
A detailed discussion on the impacts of the mass ratio and other parameters is given in \ref{sec:AppendixB}.
{\color{black} Regarding the modeling of the GW signal, we utilize the \texttt{SEOBNRv4} time-domain waveform~\cite{PhysRevD.95.044028,PhysRevD.98.084028} provided by the open-source code \texttt{PyCBC}~\cite{PyCBC}, and the phase at coalescence is set as $\varphi_c = 0$. }

\begin{figure}%%[H]
	\centering
  \includegraphics[width=.48\textwidth]{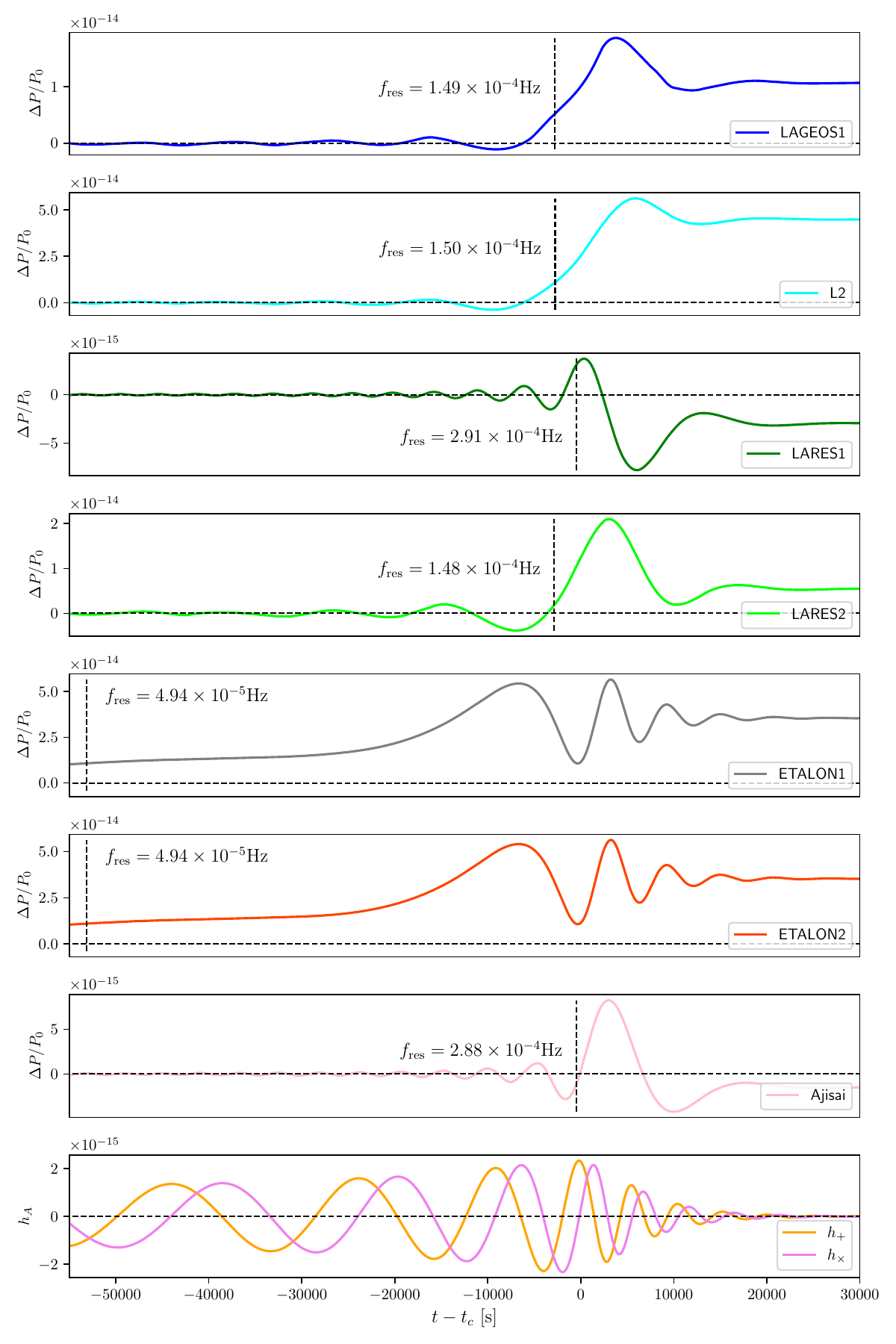}
	\caption{{ The responses of 7 laser-ranged satellites, including LAGEOS 1/2, LARES 1/2, ETALON 1/2 and Ajisai, to the same GW signal (the lowest panel) emitted by the {\bf TS}.} In each panel, the vertical line represents the time when $f_{\rm GW} = f_{\rm res}$. Note that the resonance frequencies of ETALON 1/2 are much smaller than other satellites, therefore for these two satellites, we have labeled the times corresponding to $f_{\rm res}$, rather than showing them as vertical lines. Resonances take place consecutively among these SLR missions when the chirping signal sweep across the ``frequency comb'' in the sub-mHz frequency range.}
	\label{fig:SLR_responses}
\end{figure}

Roughly speaking, the inspiral stage ends when $f_{\rm GW}$ equals the GW frequency of the innermost stable circular orbit $f_{\rm ISCO} \equiv 1/(6^{3/2}\pi M)$, where $M$ is the total mass of the source. For the case under consideration, $f_{\rm res} \approx 1.5\times 10^{-4} {\rm Hz} > f_{\rm ISCO}$, indicating that resonance happens mainly during the merger stage. 
As is shown in the top panel of Fig.~\ref{fig:SLR_responses}, for \textbf{L2}, $P$ exhibits a monotonic growth for $\sim 10^4$ s, and finally reaches a steady value with $\Delta P_{\rm fin} / P_0 = 4.374 \times 10^{-14}$.

The resonant responses of different SLR missions to the same GW signal from the \textbf{TS} are also shown in Fig.~\ref{fig:SLR_responses}. 
As expected, the secular variation $\dot{P}_{sec}$ can be either positive or negative, depending on the parameters of satellite orbits and GW sources. 
The resonance frequencies of these satellites form a ``comb'' in the sub-mHz frequency range, and resonances would take place consecutively among these SLR missions when the chirping signal sweep across the ``comb tooth''. 
Hence, the correlations among such resonant events could give a high confidence level of the detection, and may even help to investigate the physical properties of the corresponding GW sources, like the \textbf{TS}. 
To make such detection scheme attainable, the more sophisticated and important observable, that of the residual separation $\delta r(t)$, is defined and employed in the following discussions, see Fig. \ref{fig:r_residual} and \ref{fig:residual-SLRs}. 
% its post-resonance evolution will be dominated by periodic oscillations with linearly varying amplitude, see Fig. \ref{fig:r_residual} and Fig. \ref{fig:residual-SLRs}. 
% This example offers a glance to the phenomenon of resonance induced by the chirping wavetrains from SMBHBs. 

\begin{table}%%[H]
	\caption{Parameters of LAGEOS2 ({\bf L2}) and SDSSJ1430 + 2303 ({\bf TS}) in our example. }\label{tab:Parameters}
  \begin{center}
	\begin{tabular}{l|cccccc}
    \hline
    \hline
    \multirow{2}*{{\bf L2}} &
     $P_0$ & $e_0$ & $I_0$ & $\Omega_0$ & $\omega_0$  & $\varepsilon_0$ \\
     \cline{2-7}
      & $13349 {\rm s}$ & $0.0135$ & $52.64^\circ$ & $\pi / 3$ & \ $0$ \  & \ $0$ \ \\     
    \hline
    \multirow{2}*{{\bf TS}} & 
     $M_{\rm bh}$ & $z$ & $\iota$ & $\vartheta$ & $\phi$  & {\color{black} $\varphi_c$} \\
     \cline{2-7}
     & $4\times 10^7 M_{\bigodot}$ & $0.08105$ & $\pi / 6$ & $\pi / 6$ & \ $0$  \ & {\color{black} $0$} \ \\     
    \hline
    \hline 
    \end{tabular} 
  \end{center}
\end{table}

\section{GW Detection with orbital resonance}
Based on the above example, the GW-induced secular change in semi-major could only reach $\Delta a_{\rm fin}\approx \frac{2a}{3P} \Delta P_{\rm fin} \sim 10^{-7}$. Compared with the resolution of SLR distance measurements, which is at millimeter or sub-millimeter level~\cite{2019NatSR...915881C}, it seems difficult to identify such small changes in the semi-major out of uncertainties.

On the other hand, SLR is particularly superior in tracking the orbital dynamics of satellites. 
Collecting the round-trip times of laser pulses allows one to track the ``normal point'' distances over time. 
The orbital elements of the laser-ranged satellite can be derived based on such distance measurements with the help of the precise orbit determination programs, such  as \texttt{GEODYN}~\cite{GEODYN}. 
This inspired us to make use of more sophisticated observables instead of the averaged orbital elements, that of the residual normal point distance $\delta r(t)$ or the residual acceleration $\delta \vec{a}(t)$, to reveal the signatures from the GWs of coalescing SMBHBs. 
The residual distance is defined as 
\begin{equation}\label{eq:residual_r}
% \delta r(t)\equiv r(t; \boldsymbol{X}) - r(t; \rm{\hat{X}}), 
\delta r(t) \equiv r_{\rm data}(t) - r(t; \boldsymbol{X}_0) - \delta r_{\rm mod}(t),
\end{equation}
where $r_{\rm data}(t)$ is obtained from the SLR measurements, and $r(t; \boldsymbol{X}_0)$ is calculated from the initial elements $\boldsymbol{X}_0$ via Eq.(\ref{eq:Unperturbedr}). 
$\delta r_{\rm mod}(t)$ consists of the contributions of all other modeled perturbations except for GWs.
The residual acceleration is defined in the similar way $\delta \vec{a}(t)\equiv \ddot{\vec{r}}(t; \boldsymbol{X}) - \ddot{\vec{r}}(t; \boldsymbol{X}_0)-\delta \ddot{\vec{r}}_{\rm mod}(t)$. 
To make use of such data, one needs to accurately model and account for the possible gravitational and non-gravitational perturbations.
A wide variety of perturbations have been investigated in the literature, such as Earth geopotential harmonics~\cite{Ciufolini:2017srr,Ciufolini:2017lsi},  atmospheric drag~\cite{Iorio:2008sm,2016arXiv161102514P}, thermal-thrust effects~\cite{2003EAEJA.....7494L}, Solar radiation pressure, dynamic solid tide and ocean tide~\cite{1987npsg.book.....M,Lucchesi_2015}, etc.

\begin{figure}%%[H]
	\centering
    \includegraphics[width=.45\textwidth]{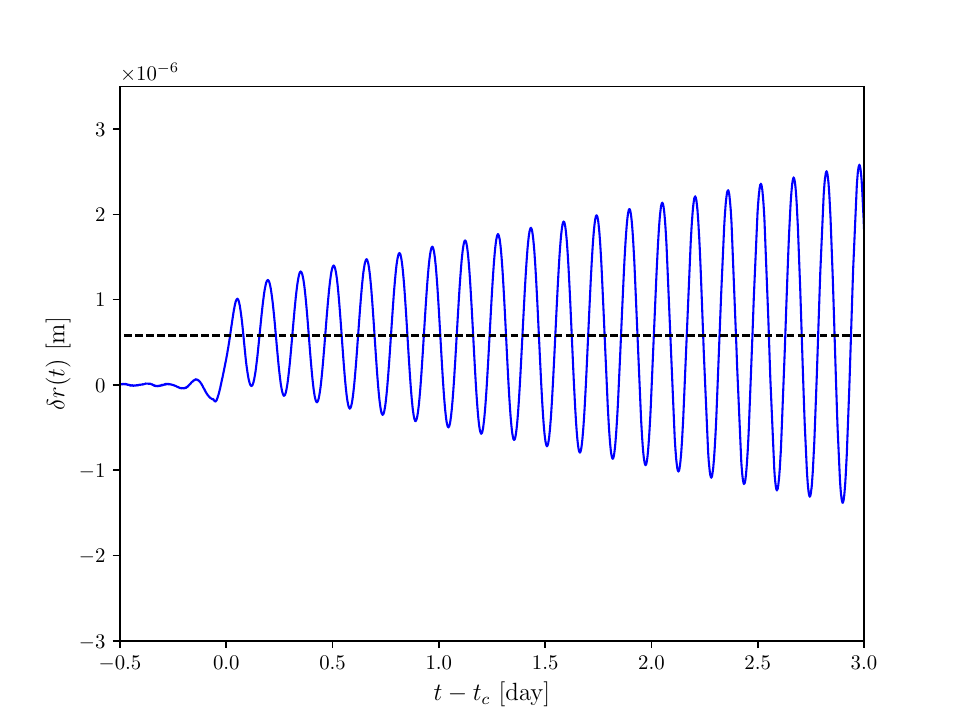} 
	\caption{{The residual distance of {\bf L2} in the most optimistic case.} The mean value of $\delta r(t)$ after resonance equals $\Delta a_{\rm fin} = 5.67 \times 10^{-7}$ m. }
	\label{fig:r_residual}
\end{figure}

\begin{figure}%%[H]
    \centering
    \includegraphics[width=.48\textwidth]{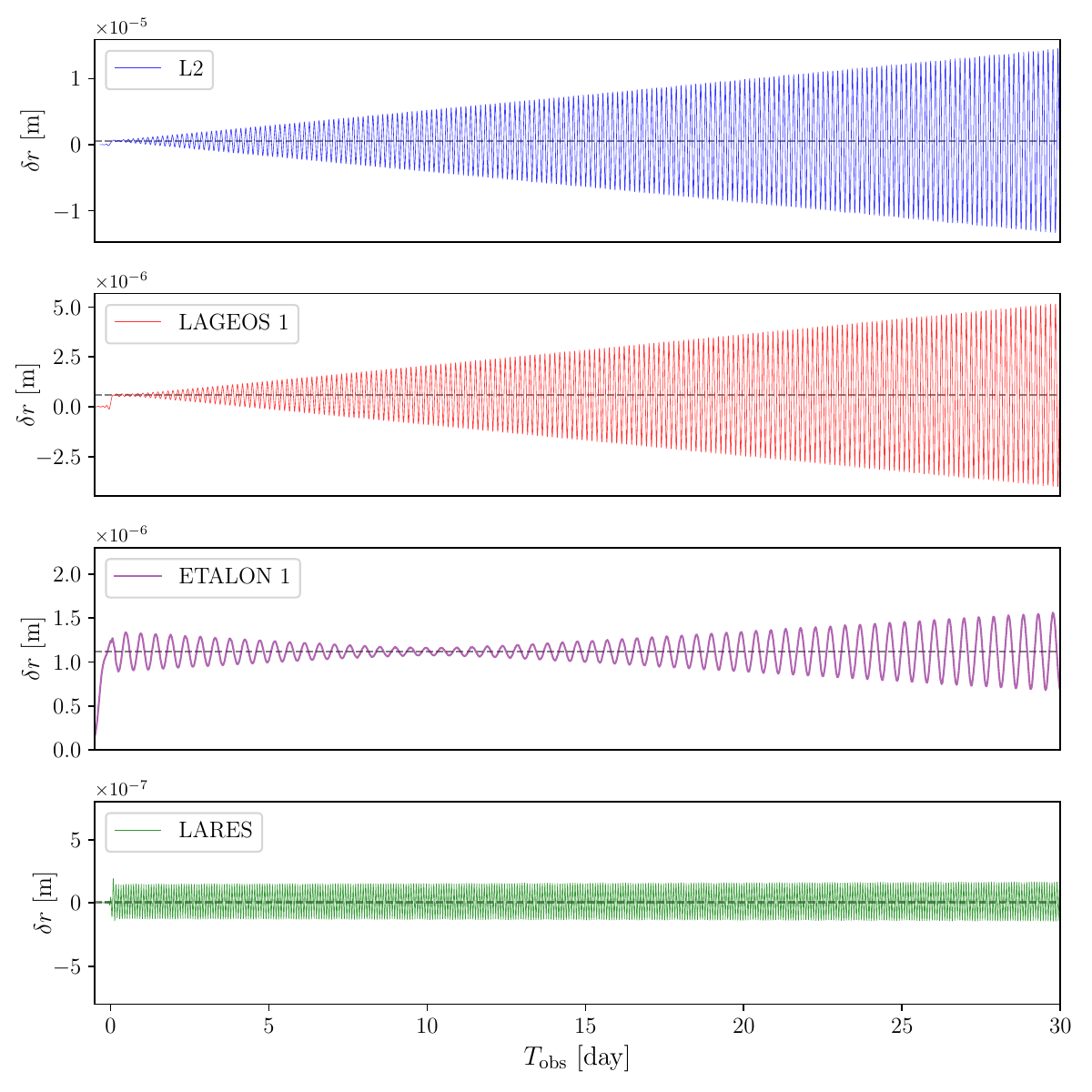}
    \caption{The long-term (30 days) distance residuals of LAGEOS 1/2, ETALON 1 and LARES 1 calculated based on their optimal responses.}
    \label{fig:residual-SLRs}
\end{figure}

With these tools, one could track the long-term dynamical evolutions of the orbits and investigate in details the differences (residuals) between the observed and modeled orbits to search for the expected signals. 
Such data analysis method is slightly different from the one used for interferometric GW detectors~\cite{Dhurandhar2022}, but has already been employed in SLR missions in a wide range of literature~\cite{Lucchesi_2015,2016arXiv161102514P,2019NatSR...915881C,Ciufolini:2012rx,Ciufolini:2013jsa}.
Considering residual accelerations, after modelling the known perturbations, the residual mean accelerations deviated away from the geodesic motion for \textbf{L2} (or LARES) are less than $1$-$2 \times 10^{-12} {\rm m/s^2}$ (or $0.5\times 10^{-12} {\rm m/s^2}$)  \cite{Ciufolini:2012rx,Ciufolini:2013jsa}. 
While, for the optimal response of {\bf L2} to {\bf TS} (see \ref{sec:AppendixB} for the determination of ``optimal parameters''),  the radial residual acceleration $\delta a_r$ will oscillate around $-2.5 \times 10^{-13} {\rm m/s^2}$ at the start of the post-resonance stage.
This order-of-magnitude estimate gives a rather optimistic evaluation of the feasibility of this new detection method.

For the convenience of SNR estimations, we use the residual distance in the following analysis and assume the ideal case that the only perturbation to the satellite orbit is from the incident GWs of SMBHBs. And, considering the expected event rate of coalescing SMBHBs \cite{PhysRevD.93.024003}, resonances in SLR measurements can be treated as individual events.  
Shown in Fig.~\ref{fig:r_residual} is the optimal response of {\bf L2} to {\bf TS} in terms of $\delta r(t)$, and in Fig.~\ref{fig:residual-SLRs} the comparison of optimal responses of different SLR missions.  
During resonance (e.g. $t \in (0.2, 0.4)$ day for {\bf L2}),  $\delta r(t)$ grows in time, and finally reaches a steady value $\overline{\delta r}_{\rm fin} = \Delta a_{\rm fin}$. Afterwards, in the post-resonance stage, the behavior of $\delta r(t)$ is in consistence with our theoretical prediction (see Appendix \ref{sec:appendixC} for the derivation)
\begin{eqnarray}\label{eq:residual_post}
    \delta r(t) 
    &\approx& \Delta a_{\rm fin}\left[ 1 - e\left(1 + \frac{\Delta e_{\rm fin} / e}{\Delta a_{\rm fin} / a}\right)\cos M \right. \nonumber \\
    & &   - \left. e\left(\frac{3\pi t}{P} - \frac{\Delta \varepsilon_{\rm fin}}{\Delta a_{\rm fin} / a}\right)\sin M + \mathcal{O}(e^2) \right],
\end{eqnarray}
where $M = 2\pi t / P + \varepsilon$ is the mean anomaly.
That is, under the long-term condition ($3\pi t / P \gg 1$), $\delta r(t)$ would oscillate around $\Delta a_{\rm fin}$ with linearly varying amplitude, and the rate of variation is proportional to $e$.
% $\delta r$ oscillates around $\overline{\delta r}_{\rm fin}$ with increasing amplitude. {\color{black} For more general cases, as can be inferred from Eq.(\ref{eq:residual_post}), the amplitude of $\delta r$ would evolve linearly in time under the long-term condition ($M \gg 1$).}
% This post-resonance residual will be absent in the case of non-resonance interactions with GWs, and 
After subtraction of the known and modeled perturbations, if the similar behaviors as in Fig.~\ref{fig:r_residual} and \ref{fig:residual-SLRs} were observed in the residuals for different SLR missions, it would indicate with high confidence that the GW-induced resonance as the cause. 
Another conclusion which can be drawn from Fig. \ref{fig:residual-SLRs} is that the eccentricity of satellite orbit plays an important role in the post-resonance evolution, as is predicted by Eq. (\ref{eq:residual_post}). Indeed, the growth rate of $\delta r$ for {\bf L2} ($e = 0.0135$) is relatively large compared to, for example, LAGEOS 1 ($e = 0.0045$), ETALON 1 ($e = 0.001$) and LARES 1 ($e = 0.001$).
% is nothing but the difference between two unperturbed elliptic Kepler orbits with slightly different orbital elements. 
% Trivial as it may seem, this pattern is the consequence of secular evolution, and it is absent in the case of non-resonance interactions with GW. 
% Next, we will estimate whether the growth of $\delta r$ in the resonance stage, together with this pro-resonance pattern, can be used to detect GW signal.  

To extract the signal of GW, the residual data should be analyzed with methods such as matched filtering. Based on long-term data tracking, the SNR (dubbed $\rho$) for the optimal response of {\bf L2} can be approximated as 
\begin{equation}\label{eq:DefSNR}
\rho^2 = 4\int_{0}^{\infty} \frac{|\delta \tilde{r}(f)|^2 \ df}{S_n(f)}\approx \frac{\Delta a_{\rm fin}^2}{\sigma^2 t_s}\left(T_{\rm obs} + \frac{3\pi^2 e^2}{2P^2}T^3_{\rm obs}\right),
\end{equation}
where $\delta \tilde{r}(f)$ is the Fourier transform of $\delta r(t)$, and $S_n(f)$ represents the one-sided noise power spectral density of SLR. 
The resonance stage lasts less than 1 day and contributes a rather small fraction to the total SNR. 
Whereas, in the post-resonance stage, $\delta r(t)$ oscillates with growing amplitude and results in the above polynomial SNR on the observation time $T_{\rm obs}$ (see Appendix \ref{sec:appendixC}).
$S_n(f)$ depends on the uncertainty $\sigma$ of normal point measurement{\color{black}. Ref.~\cite{Blas:2021mqw} predicted that the precision of SLR coincides with that of LLR, and the latter will have an order-of-magnitude improvement in the next decade (which might require the installation of new retroreflectors~\cite{Murphy:2013qya}). Following their assumption, we consider two values of $\sigma$ in this paper:}
\begin{enumerate}
    \item [a.] current precision: $\sigma = 3$ mm, $50,000$ normal point measurements per year;
    \item [b.] improved precision: $\sigma = 0.3$ mm, $200,000$ measurements per year.
\end{enumerate}

\begin{figure}%%[H]
	\centering
    \includegraphics[width=.45\textwidth]{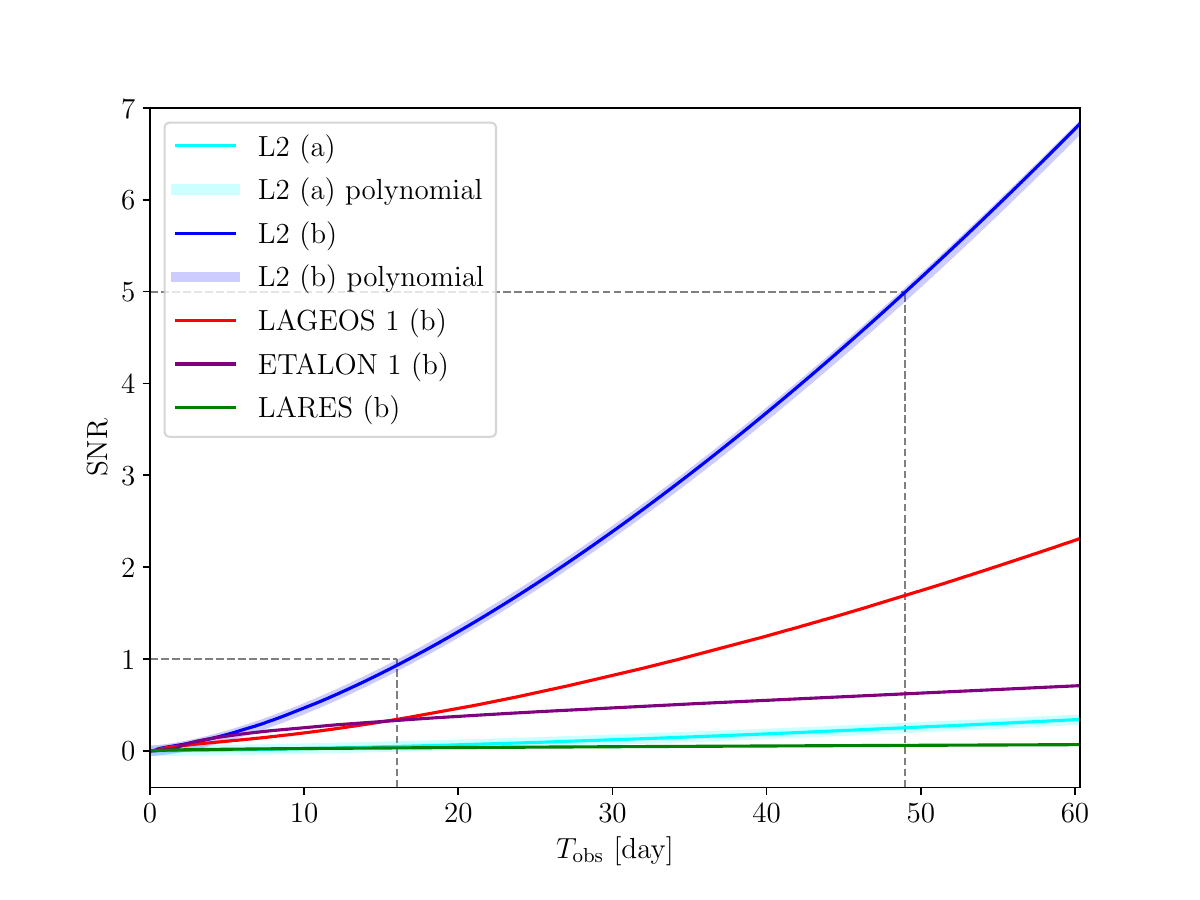}
	\caption{The SNRs of the most optimistic responses to {\bf TS} for different satellites. The thin curves are obtained by numerical calculation, while the thick ones represent the polynomial function Eq.(\ref{eq:DefSNR}).}
	\label{fig:SNR_residual_longlong}
\end{figure}

In Fig.~\ref{fig:SNR_residual_longlong} we plot the SNR against $T_{\rm obs}$ for different missions under different laser ranging precision.
% obtained by means of numerical calculations and the power-law fitting functions. 
As is shown, for \textbf{L2}, SNR $>$ 1 can be achieved when $T_{\rm obs} >$ 124 days (current precision a) or 17 days (improved precision b). 
Once we set a realistic threshold for GW detection to SNR = 5, $T_{\rm obs} >$ 356 days are required for precision a and 49 days for precision b, which turns out to be practical and workable.

\begin{figure}%%[H]
	\centering
        \includegraphics[width=.45\textwidth]{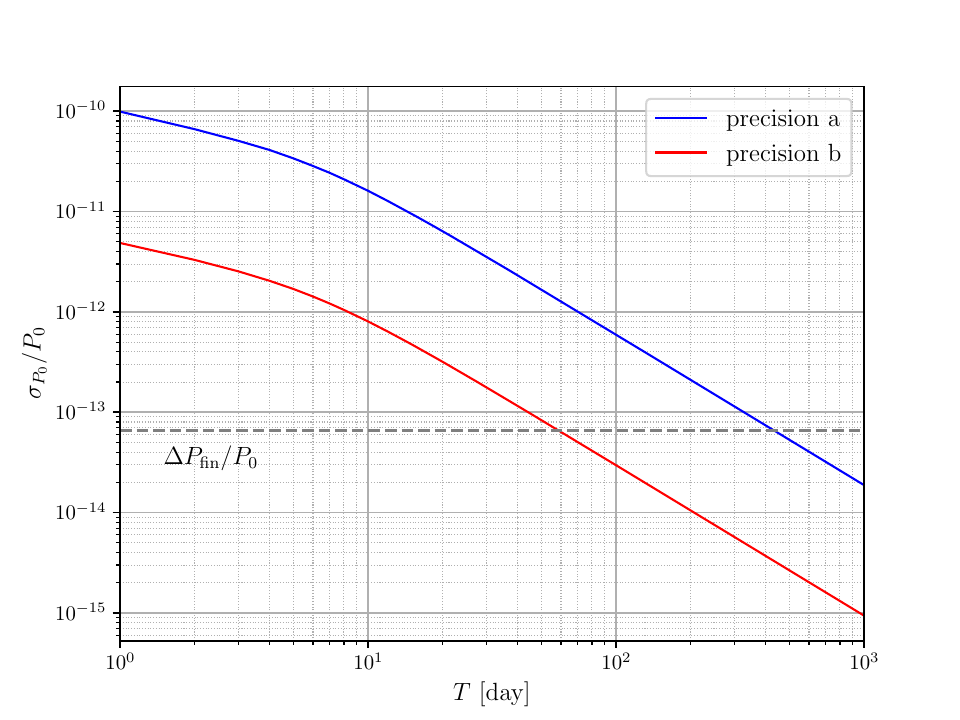}
	\caption{The relationship between the error of theeorbital period $\sigma_{P_0}$ and the in-orbit time $T$ used for calculating $P_0$. The blue curve and the red curve represents the results based on precision a and precision b, respectively. For comparison, the optimal response of \textbf{L2} to \textbf{TS} is also plotted with grey dashed line in the same figure.}
	\label{fig:precision_vs_signal}
\end{figure}

{\color{black} 
In the derivation of Eq.(\ref{eq:DefSNR}), it is implicitly assumed that we have perfect knowledge of the unperturbed orbital period, thus $S_n(f)$ does not account for the contribution of the prior uncertainty of $P_0$ (dubbed $\sigma_{P_0}$ hereafter). This can only be achieved with infinitely long in-orbit time used for the calibration of orbital elements. Therefore, it is necessary to examine whether the impact of $\sigma_{P_0}$ can be safely neglected compared to the effect of passing GWs, provided that $P_0$ is determined from a reasonable number of SLR measurements. To estimate $\sigma_{P_0}$, we employ the  Fisher information matrix (FIM) formalism presented in Ref.~\cite{PhysRevD.105.064021}, which converts the uncertainties of laser ranging to those of orbital elements. Within a given time $T$ (which does not have to coincide with $T_{\rm obs}$), 
\begin{equation}
    F_{ij} = \frac{1}{\sigma^2} \sum_t^{T} \partial_i r(t) \partial_j r(t) , \quad \sigma_i = \sqrt{\left(F^{-1}\right)_{ii}},
\end{equation}
where $i$ represents the parameters relevant to $r(t)$, i.e. $i = P, e, \varepsilon$. The $\sigma_{P_0} - T$ relationship for \textbf{L2}, as well as the target signal (e.g. the optimal response of \textbf{L2} to \textbf{TS}) are plotted in Fig.~\ref{fig:precision_vs_signal}. It is clearly shown that, under precision b, the prerequisite ($\sigma_{P_0} \ll \Delta P$) for our SNR calculation is fully satisfied with $T \sim 10^3$ days . This is achievable for missions like LAGEOS which have been operating for decades. While for precision a, the SLR data collected over a period of  $\sim 10^3$ days only yield a $\sigma_{P_0}$ comparable to the signal, necessitating a longer time of calibration, or the signal should be stronger by one or more orders (such as the examples given in the following analysis).} 

{\color{black}In addition, it should be noted that although the prior uncertainty of $P_0$ has been considered,  this is still an ideal scenario in the sense that other gravitational or non-gravitational perturbations are well modeled and subtracted from the data. Given the specific models of these perturbing factors to the SLR missions, the error analysis would become more realistic, while this is beyond the scope of our work and belongs to a separate research topic.}

The magnitudes of resonant responses also depends on the properties of GW sources, especially the masses and redshifts (see Appendix \ref{sec:MandZ}), we then go beyond {\bf TS} and look for more promising candidates. 
{\color{black} The nearest SMBHB system reported so far is located in NGC 7277, with redshift $0.006$ and component masses $1.54 \times 10^8 M_{\bigodot}$ and $6.3 \times 10^6 M_{\bigodot}$~\cite{refId0}. Although this system itself is not an imminent merging one, the existence of SMBHB at redshift around $z \sim 0.01$ can not be ruled out. }
Suppose that {\bf L2}, as the representative, is in resonant interaction with the GW from a SMBHB with redshift $z = 0.01$ and equal component masses $M_{\rm bh} = 5.9 \times 10^7 M_{\bigodot}$,
% (corresponding to the peak of the red curve in Fig.~\ref{fig:dP_vs_m}), 
in the most optimistic case,
SNR = 5 can be achieved after an observation time of 68 days (current precision a) or 9 days (improved precision b). 
Furthermore, for an imaginary SMBHB at $z = 0.001$, by only taking the data within resonance stage into consideration, the most optimistic SNR could reach 0.15 (current precision a) or 3 (improved precision b). 
To the far end, for mergers of {\bf TS}-like SMBHBs at $z = 0.1$, data sets of 410 days (current precision a) or 56 days (improved precision b) are needed to achieve ${\rm SNR} > 5$. 
The above analysis indicates that this new method could give tentative detections of the violent mergers of SMBHBs within the reach of $z \sim 0.1$. 
Moreover, as expected, when considering the joint detection by all SLR missions in operation, both the total SNR and the confidence level could be further improved. 

\section{Concluding remarks}
In this work, we have investigated the feasibility of the detection scheme for GWs from coalescing SMBHBs, through their resonant interactions with the laser-ranged satellites. 
% Both numerical and analytical methods were employed to analyze the resonant response of satellite orbits to the incident GWs. 
The observable of residual distance or residual acceleration are introduced to make the detection attainable. 
The SNR of measuring the resonance-induced characteristic signals in the residual distances and the dependence on relevant parameters of GW sources and orbiters are analyzed.
It turns out that, before the launches of the space-borne antennas, SLR may be the only ready-to-use approach of probing coalescing SMBHBs in the sub-mHz range.
% which may even provide us opportunities for multi-messenger observations.

Among the promising candidates, we take SMBHB SDSSJ1430+2303 ({\bf TS}) as the representative example, which is expected to merge within 3 years.  
For LAGEOS 2, we discussed the SNR of detecting the merger of \textbf{TS} for two sets of ranging precision. 
In the optimistic case, SNR $>$ 1 can be achieved when $T_{\rm obs} >$ 124 days for the current precision or 17 days for the future improved precision. 
For the threshold SNR = 5, 356 days is required for the current precision, and 356 days for improved precision. 
These results are generated to similar sources and SLR missions. For \textbf{TS} like candidates at redshift $z=0.1$, ${\rm SNR}=5$  requires less than two years data for the current precision and a few months data for the improved precision. These are workable for data processing of SLR missions.
Moreover, as the chirping waves sweep across the``frequency comb'' in the sub-mHz range, the possible joint detection by the multiple laser-ranged satellites could further improve the total SNR and the detection confidence. 

To summarize, SLR missions with the resonant detection scheme could fulfill the requirement of a tentative SMBHB probe that within the reach of $z\sim 1$. 
Not just future-oriented, the re-analysis of the archived data from the past decades with our method is also worthwhile.   
At last but not least, the prospect of our method also improves with the understandings of the total orbital perturbations for SLR missions.

\begin{acknowledgement}
    This work is supported by the National
    Key Research and Development Program of China,
    No. 2020YFC2200601 and No. 2021YFC2201901.
\end{acknowledgement}

%%%%%%%%%%%%%%%%%%%%%%%%%%%%%%%%%%%%%%%%%%%%%%%%%%%%%%%
%%% Appendix sections. ??????, ????
%%%%%%%%%%%%%%%%%%%%%%%%%%%%%%%%%%%%%%%%%%%%%%%%%%%%%%%
% \begin{appendix}
\section*{Appendix}
\appendix 
  \section{Derivation of the simplified analytical solution \label{sec:appendixA}}
  An analytical solution can be obtained under the conditions that the test binary’s orbit is near circular ($e \ll 1$) and the incident GW is modeled as a monochromatic wave:
  \begin{eqnarray}
      h_+ &=& H_+ \cos(2\pi f_{\rm GW} t + \varphi_{\rm GW}), \nonumber \\
      h_\times &=& H_\times \sin(2\pi f_{\rm GW} t + \varphi_{\rm GW}),
  \end{eqnarray}
  where $f_{\rm GW}$ and $\varphi_{\rm GW}$ are the redshifted frequency and initial phase of GW, respectively. The amplitudes $H_A (A = +, \times)$ can be expressed in terms of the redshifted chirp mass $\mathcal{M}_c$, luminosity distance $d_L$ and inclination $\iota$ as
  \begin{eqnarray}
      H_+ &=& \frac{4 \mathcal{M}_c^{5/3}(\pi f_{\rm GW})^{2/3}}{d_L} \frac{1 + \cos^2 \iota}{2}, \nonumber \\
      H_\times &=& \frac{4 \mathcal{M}_c^{5/3}(\pi f_{\rm GW})^{2/3}}{d_L} \cos \iota.
  \end{eqnarray}
 It will be demonstrated that the resonance is strongest in the face-on case. In this scenario, the inclination angle $\iota$ equals zero, thus the polarization angle $\psi_{\rm P}$ and the initial phase of GW $\varphi_{\rm GW}$ are complete degenerate. Hence for simplicity we will omit the dependence on $\psi_{\rm P}$ in the subsequent analysis.  
 
 In the case of small orbital variation, we can expand $\boldsymbol{X}$ in powers of $h_A$ as $\boldsymbol{X} = \boldsymbol{X}_0 + \boldsymbol{X}_1(t; h_A) + \mathcal{O}(h_A^2)$, insert this expansion into Eq.({\color{black}6}) in the body of the paper, and keep only the linear order.  
  Taking the orbital period as an example, it follows that
  \begin{equation}\label{eq:LinearPerturbedP}
      % \dot{P} = \frac{3P_0^2\gamma_0}{4\pi}\left[\frac{e_0}{\gamma_0^2}\frac{r}{a_0}\sin\psi e_{ij}^A \hat{r}^i \hat{r}^j + e_{ij}^A \hat{\theta}^i \hat{r}^j \right]\ddot{h}_A
      \dot{P} = T^A_P\left[\boldsymbol{X}_0, \psi(t; \boldsymbol{X}_0), \boldsymbol{\hat{n}}_{\rm GW}\right] \ddot{h}_A(\boldsymbol{\hat{n}}_{\rm GW}, t).
  \end{equation}
  In the rest of this appendix we will drop the subscript ``0'' for brevity, and one should keep in mind that $\boldsymbol{X}$ in the r.h.s. stand for $\boldsymbol{X}_0$. 
  
  The most cumbersome parts of $T^A_P$ are $e_{ij}^A\hat{r}^i\hat{r}^j$ and $e_{ij}^A\hat{r}^i\hat{\theta}^j$, which take the following forms in the frame of test binary:
  \begin{eqnarray}
      e_{ij}^A\hat{r}^i\hat{r}^j &=& C^A_r \cos 2\theta + S^A_r \sin 2\theta + K^A, \nonumber \\
      e_{ij}^A\hat{r}^i\hat{\theta}^j &=& C^A_\theta \cos 2\theta + S^A_\theta \sin 2\theta,
  \end{eqnarray}
  with $\theta \equiv \omega + \psi$, and
  \begin{eqnarray}
      C^+_r &=& \frac{1}{2}\left(\cos^2\varphi \cos^2 I - \sin^2 \vartheta \sin^2 I - \cos^2 \vartheta \sin^2 \varphi \cos^2 I \right.  \nonumber \\ 
      & &  \ + \left. \frac{1}{2}\sin 2\vartheta \sin \varphi \sin 2I - \sin^2 \varphi + \cos^2 \vartheta \cos^2 \varphi\right), \nonumber \\ 
      S^+_r &=& -\frac{1}{2}\left(\sin 2\vartheta \cos \varphi \sin I + \sin^2 \vartheta \sin 2\varphi \cos I\right), \nonumber \\
      K^+ &=& \frac{1}{2}\left(-\cos^2\varphi \cos^2 I + \sin^2 \vartheta \sin^2 I + \cos^2 \vartheta \sin^2 \varphi \cos^2 I \right.  \nonumber \\
      & &  \ - \left. \frac{1}{2}\sin 2\vartheta \sin \varphi \sin 2I - \sin^2 \varphi + \cos^2 \vartheta \cos^2 \varphi\right), \nonumber 
  \end{eqnarray}
  \begin{eqnarray}\label{eq:Coefficients}
      C^\times_r &=& \frac{1}{2}\left(-\cos\vartheta \sin 2\varphi \cos^2 I + \sin \vartheta \cos \varphi \sin 2I \right. \nonumber \\
      & & \ - \left. \cos \vartheta \sin 2\varphi\right), \nonumber \\
      S^\times_r &=& \cos\vartheta \cos 2\varphi \cos I + \sin \vartheta \sin \varphi \sin I, \nonumber \\
      K^\times &=& \frac{1}{2}\left(-\cos\vartheta \sin 2\varphi \sin^2 I - \sin \vartheta \cos \varphi \sin 2I \right), \nonumber \\
      C^+_\theta &=& -\frac{1}{2}\sin 2\vartheta \cos\varphi \sin I + \frac{1}{2}(1 + \cos^2 \vartheta) \sin 2\varphi \cos I, \nonumber \\
      S^+_\theta &=& \frac{1}{2}\sin^2 \varphi - \frac{1}{2}\cos^2\varphi \cos^2 I - \frac{1}{2}\cos^2 \vartheta \cos^2 \varphi  \nonumber \\
          & &  \  + \frac{1}{2}\left(\sin \vartheta \sin I - \cos \vartheta \sin \varphi \cos I \right)^2, \nonumber \\ 
      C^\times_\theta &=& \cos\vartheta \cos 2\varphi \cos I + \sin \vartheta \sin \varphi \sin I, \nonumber \\
      S^\times_\theta &=& -\frac{1}{2}\left[\sin \vartheta \cos \varphi \sin 2I +\cos \vartheta \sin 2\varphi (1+\cos^2 I)\right], \nonumber \\
  \end{eqnarray}
  where we have defined $\varphi \equiv \phi - \Omega$ for convenience.
  
  For elliptic orbits, the explicit time-dependence of Kepler motion can be expressed in terms of the Hansen coefficients~\cite{HansenCoef,HansenCoefBessel}:
  \begin{eqnarray}
      \left(\frac{r}{a}\right)^n \sin m\psi &=& \sum_{p = -\infty}^{+\infty}X_p^{nm}(e)\sin p M, \nonumber \\
      \left(\frac{r}{a}\right)^n \cos m\psi &=& \sum_{p = -\infty}^{+\infty}X_p^{nm}(e)\cos p M,
  \end{eqnarray}
  where $M$ is the mean anomaly $M = 2\pi t / P + \varepsilon$. As a result, up to the linear order of $e$, we have
  \begin{eqnarray}\label{eq:dP_eccentric}
      \dot{P} = &-&3\pi \gamma \alpha^2 H_A\cos\left(\frac{2 \pi \alpha}{P} t + \varphi_{\rm GW} - \delta_{A \times}\frac{\pi}{2}\right) \nonumber \\
      &\times& \left[\sqrt{G_{1A}^{2} + G_{2A}^{2}} \sin \left(2M + \arctan \frac{G_{1A}}{G_{2A}}\right)\right. \nonumber \\
      & & \ + \ e\sqrt{\left(F_{4A} - 2G_{1A}\right)^2 + \left(F_{2A} - 2G_{2A}\right)^2} \nonumber \\
      & & \quad \ \times\sin\left(M + \arctan\frac{F_{4A} - 2G_{1A}}{F_{2A} - 2G_{2A}}\right) \nonumber \\
      & & \ + \ e\sqrt{\left(F_{3A} + 2G_{1A}\right)^2 + \left(F_{1A} + 2G_{2A}\right)^2} \nonumber \\
      & & \quad \ \times\sin\left(3M + \arctan\frac{F_{3A} + 2G_{1A}}{F_{1A} + 2G_{2A}}\right) \nonumber \\
      & & \ + \left. \mathcal{O}(e^2)\right].
  \end{eqnarray}
  In the first line, $\alpha \equiv f_{\rm GW} P$ denotes the ratio between GW frequency and the orbital frequency of the test binary, and $\delta_{A \times} = 1$ if $A = \times$ or $0$ if $A = +$. The factors $F_{iA}$ and $G_{iA}$ are combinations of $C^A_r$, $C^A_\theta$, $S^A_r$ and $S^A_\theta$ defined as
  \begin{eqnarray}
      F_{1A} &=& \frac{1}{2\gamma^2}\left(C^A_r \cos 2\omega + S^A_r \sin 2\omega \right), \nonumber \\
      F_{2A} &=& \frac{1}{2\gamma^2}\left(-C^A_r \cos 2\omega - S^A_r \sin 2\omega  + 2K^A\right), \nonumber \\
      F_{3A} &=& \frac{1}{2\gamma^2}\left(C^A_r \sin 2\omega - S^A_r \cos 2\omega \right), \nonumber \\
      F_{4A} &=& \frac{1}{2\gamma^2}\left(-C^A_r \sin 2\omega + S^A_r \cos 2\omega \right), \nonumber \\
      G_{1A} &=& C^A_\theta \cos 2\omega + S^A_\theta \sin 2\omega, \nonumber \\ 
      G_{2A} &=& -C^A_\theta \sin 2\omega + S^A_\theta \cos 2\omega. 
  \end{eqnarray}
  
  By integrating Eq.(\ref{eq:dP_eccentric}), it is straightforward to show that $P$ has a secular and linear evolution when $\alpha = 2$, which is of order $\mathcal{O}(e^0)$, and we will refer to $\alpha = 2$ as the ``main'' resonance frequency. Besides, two ``secondary'' resonances of order $\mathcal{O}(e^1)$ occur at $\alpha = 1$ and $3$, indicating that the resonant responses of eccentric binaries are more complicated than the circular ones. 
  
  Within the context of our discussion, $e \sim 10^{-3} - 10^{-2}$, thus for the simplified analytical solution we will mainly focus on the ``main'' resonance frequency $f_{\rm GW} = f_{\rm res} = 2 / P$. The secular evolution rate of $P$, defined as $\dot{P}$ averaged over one revolution, reads
  \begin{eqnarray}\label{eq:dP_eccentric_resonance}
      \dot{P}_{\rm sec} &=& 6\pi \gamma H_A\sqrt{G_{1A}^2 + G_{2A}^2} \nonumber \\ 
      &\times& \sin\left(\varphi_{\rm GW} - 2\varepsilon - \tan^{-1} \frac{G_{1A}}{G_{2A}} - \delta_{A\times}\frac{\pi}{2}\right).
  \end{eqnarray}
  Eq.(\ref{eq:dP_eccentric_resonance}) can be further simplified if we consider the case where the orbital planes of the source binary and the test binary are face-on i.e. $\{\vartheta = I, \phi = \Omega - \pi /2, \iota = 0\}$, which gives 
  \begin{equation}\label{eq:SimplifiedSecularEvolutionMax}
      \dot{P}_{\rm sec} = 12\pi \gamma H \sin(\varphi_{\rm GW} - 2\omega - 2\varepsilon),
  \end{equation}
  where $H \equiv 4 \mathcal{M}_c^{5/3} (\pi f_{\rm GW})^{2/3} d_L^{-1}$. For circular orbits, $\omega$ and $\varepsilon$ become ill-defined. This issue is usually solved by introducing the combination $\xi = \varepsilon + \omega$, which, for a stable circular orbit, stands for the angle from the ascending node to the initial position. Therefore
  \begin{equation}
      \dot{P}_{\rm sec} = 12\pi H \sin(\varphi_{\rm GW} - 2\xi).
  \end{equation}

  \section{A detailed discussion on the parameter space and the determination of optimal parameters}\label{sec:AppendixB}

    \begin{figure}%[H]
      \centering
      \includegraphics[width=0.44\textwidth]{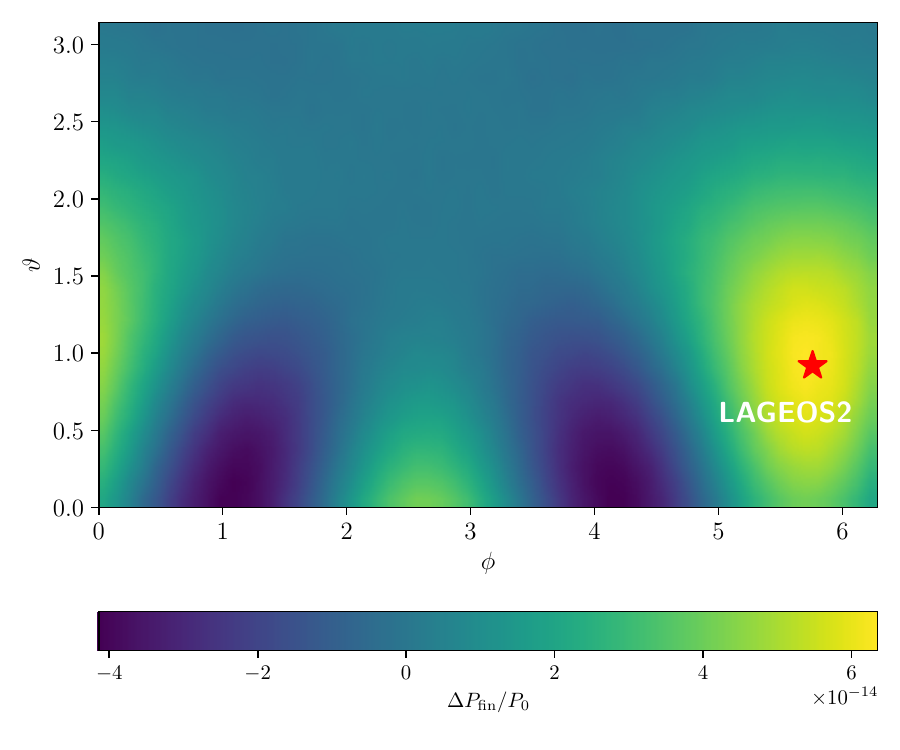}
      \caption{The dependence of  $\Delta P_{\rm fin} / P_0$ on $(\vartheta, \phi)$. The normal vector of {\bf L2}'s orbit $((\vartheta, \phi) = (0.919, 5.760))$ is marked as a red star.}
      \label{fig:dP_meshplot}
  \end{figure}
  
  The orbital resonance of the satellite depends on several parameters, including the position, orientation, redshift, and component masses of the GW source, and the initial orbital elements of the test binary. For monochromatic sources, the relationship between orbital resonance and aforementioned parameters are well manifested by the formulae deduced in~Appendix \ref{sec:appendixA}. As for chirp signal, to illustrate the impacts of these parameters, and look for the ``optimal parameters'' that maximize the effect of resonance, we vary some of their values while keeping others the same as Tab.~{\color{black}1}, and calculate the evolution of $P$ numerically. For brevity, we will leave out the subscript ``0'' of $\boldsymbol{X}_0$ in the rest of this appendix.
  
  \subsection{the celestial coordinate $(\vartheta, \phi)$ of GW source}
  The role of inclination angle $\iota$ in GW amplitude is quite straightforward. To seek for the celestial position of source which leads to maximum resonance, we set $\iota = 0$, vary $(\vartheta, \phi)$ and calculate $\Delta P_{\rm fin} / P_0$ numerically. The result is visualized in Fig.~\ref{fig:dP_meshplot}, where the normal vector of ${\bf L2}$ orbit is marked as a red star. As is shown, maximum resonance occurs when the source binary and the test binary are face-on.
  
  \subsection{$\omega$ of the test binary and {\color{black} $\varphi_{c}$} of the source binary}
\begin{figure}%[H]
  \centering
  \includegraphics[width=0.4\textwidth]{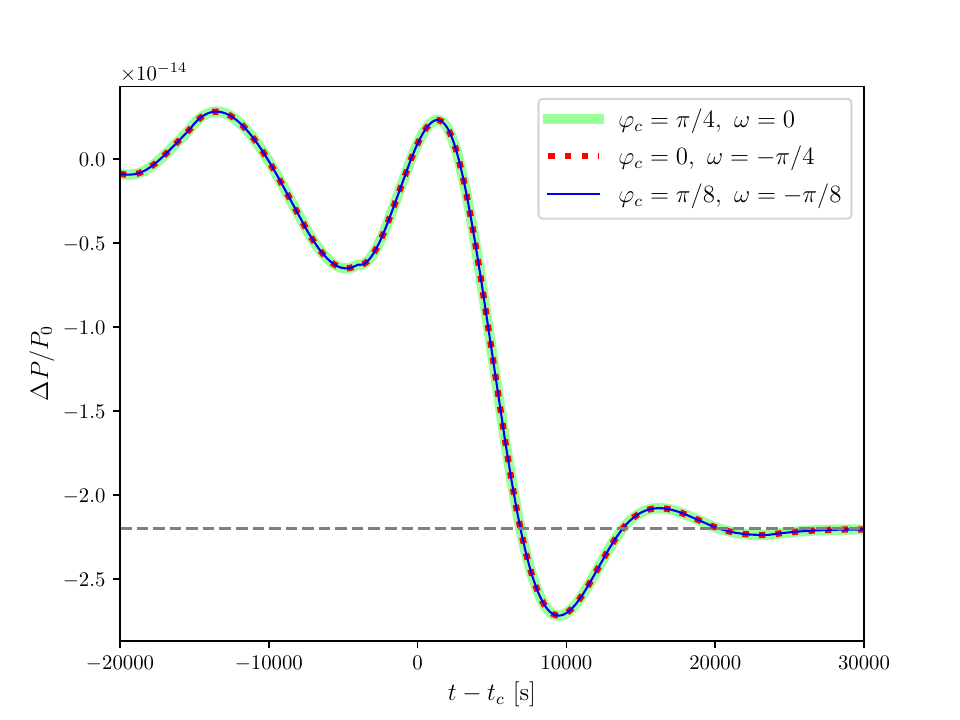} \\
  \includegraphics[width=.42\textwidth]{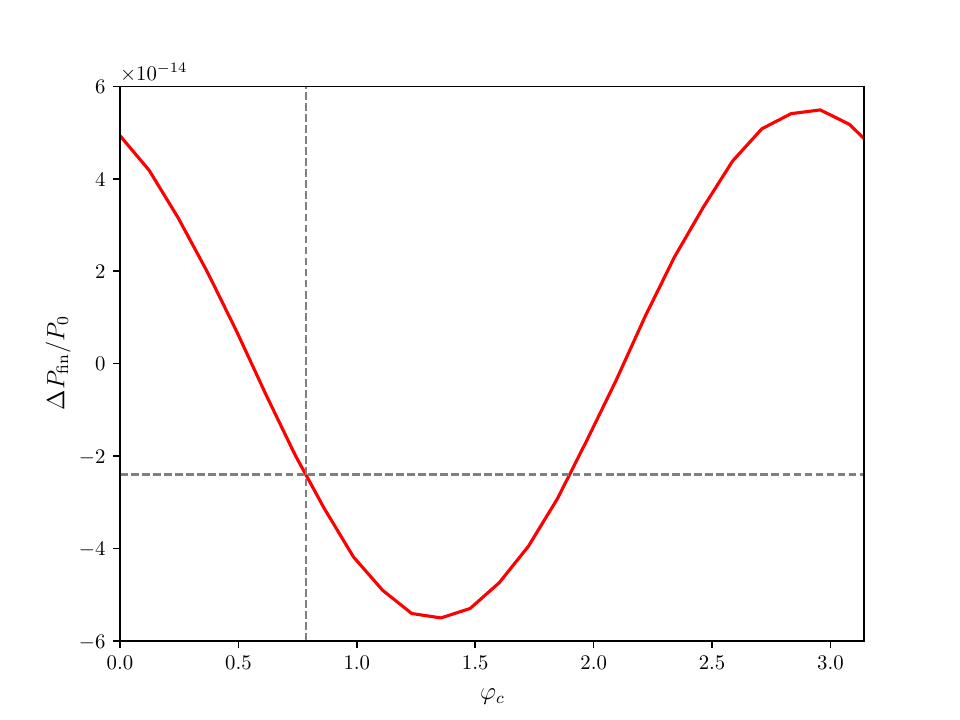}
  \caption{{\color{black}Upper panel: $\Delta P / P_0$ under 3 equivalent sets of $(\omega, \varphi_{c})$. Lower panel: the dependence of $\Delta P_{\rm fin} / P_0$ on $\varphi_{c}$ with $\omega$ fixed to 0. The first set in the upper panel is marked with dashed lines.}}
  \label{fig:omega_and_phi}
  \end{figure}

  Eq.(\ref{eq:SimplifiedSecularEvolutionMax}) indicates that the effects of $\omega$, $\varepsilon$ and $\varphi_{\rm GW}$ are degenerate, and $\dot{P}_{\rm sec}$ depends on the combination $\varphi_{\rm GW} - 2\omega$ (since we have set the initial condition $\varepsilon = 0$). This relationship also holds for chirping signals, {\color{black}only that $\varphi_{\rm GW} - 2\omega$ should be replaced by $2(\varphi_{c} - \omega)$, for $\varphi_{c}$ is defined as the source's orbital phase at coalescence. To prove this, we vary $(\omega, \varphi_{c})$ and keep their difference invariant. Shown in the upper panel of Fig.~\ref{fig:omega_and_phi} are the results of 3 equivalent combinations, and other parameters take the values in Tab.~1.} Obviously, the results of these combinations are indistinguishable. 
  
  Furthermore, we investigate the dependence of $\Delta P_{\rm fin} / P_0$ on $2 (\varphi_{c} - \omega)$, which is equivalent to varying $\varphi_{c}$ and keeping $\omega = 0$. It can be seen from the lower panel of Fig.~\ref{fig:omega_and_phi} that $\Delta P_{\rm fin} / P_0$ acts like a sinusoidal function of $2\varphi_{c}$. In the following we will denote the values of $\{\vartheta, \phi, \iota, \omega, \varphi_{c}\}$ which maximize $\Delta P_{\rm fin} / P_0$ as the Optimal Parameters ({\bf OP} hereafter).

  \subsection{the semi-major axis $a$ of test binary}\label{sec:HeightofOrbit}

\begin{figure}%[H]
      \centering
      \includegraphics[width=.43\textwidth]{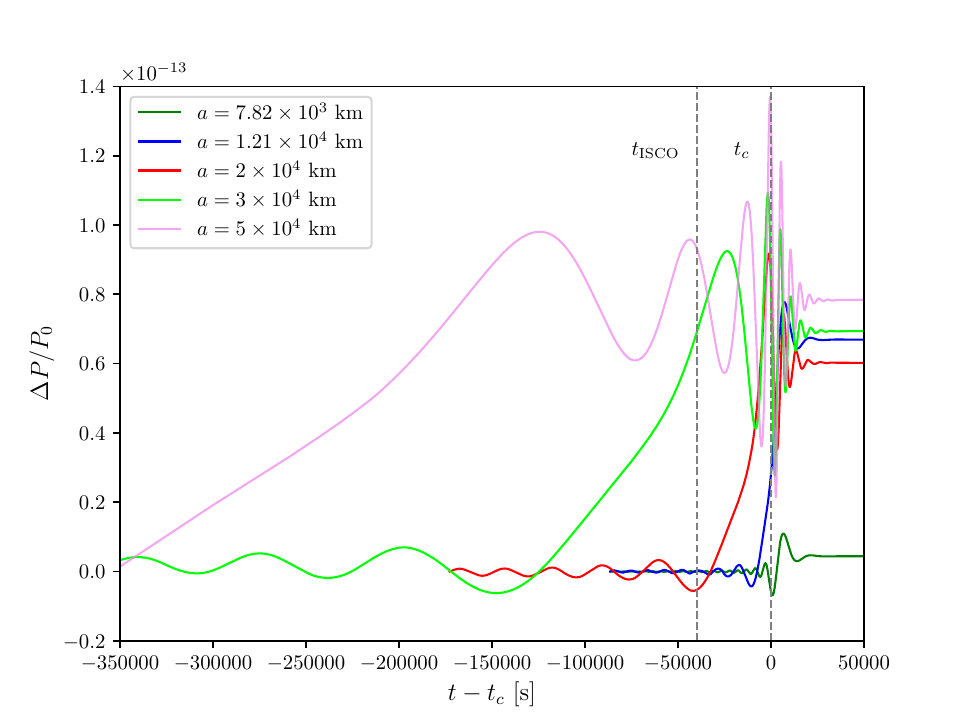}
      \caption{{The relative variations of $P$ for $a = \{0.782, 1.21, 2, 3, 4\}$ \\ $ \times 10^4$ km. $t_{\rm ISCO}$ and $t_c$ are shown with dashed lines to roughly divide the incident GW signal into different stages.}}
      \label{fig:dP_vs_a}
  \end{figure}

  We consider four values of the semi-major axis: \\$a = \{0.782, 1.21, 2, 3, 4\} \times 10^4$ km. The 1st and 2nd of them correspond to the configurations of {\bf L2} and LARES 1. Besides, to illustrate the impact of $a$ on the same basis, for each value of $a$ we iterate over $\{\vartheta, \phi, \iota, \omega, \varphi_{c}\}$, and then set them to the {\bf OP}. Given the total mass of the test binary (which is approximately the mass of Earth), the orbital frequency, and hence the resonance frequency, are totally determined by $a$. Consequently, for different values of $a$, resonance takes place at difference stages of GW (see Fig.~\ref{fig:dP_vs_a}):
  
       (1) \textbf{Inspiral} (e.g. $a = 4 \times 10^4$ km): $f_{\rm GW}$ increases slowly, thus resonance can last for a relatively long time ($\sim 10^5$ s). The resulting $\Delta P_{\rm fin} / P_0$ is the largest among all the $a$ values in consideration; 
       
       (2) \textbf{Merger}  (e.g. $a = 1.21 \times 10^4$ km, $2 \times 10^4$ km, $3 \times 10^4$ km): The duration of resonance is shorter than case (1), while $h_A$ and $\ddot{h}_A$ of this stage get much larger, thus the effect of resonance is only slightly weaker than case (1);
       
       (3) \textbf{Coalescence and ring down} (e.g. $a = 0.782 \times 10^4$ km): In this extreme situation, resonance can only last for a very short duration, leading to a much smaller $\Delta P_{\rm fin} / P_0$.
  
  % For most of the cases, the values of $\Delta P_{\rm fin} / P_0$ are at the same order, however, the absolute changes in $a$ are considerably discrepant, i.e. 
  For the variation of $a$, when $a = \{0.782, 1.21, 2, 3, 4\} \times 10^4$ km, $\Delta a_{\rm fin} = \{0.23, 5.67, 8.03, 13.89, 20.91\} \times 10^{-7}$ m.

  \subsection{the component masses $M_{\rm bh}$, redshift $z$ and mass ratio $q$ of SMBHB}\label{sec:MandZ}
  
\begin{figure}%[H]
      \centering
      \includegraphics[width=.42\textwidth]{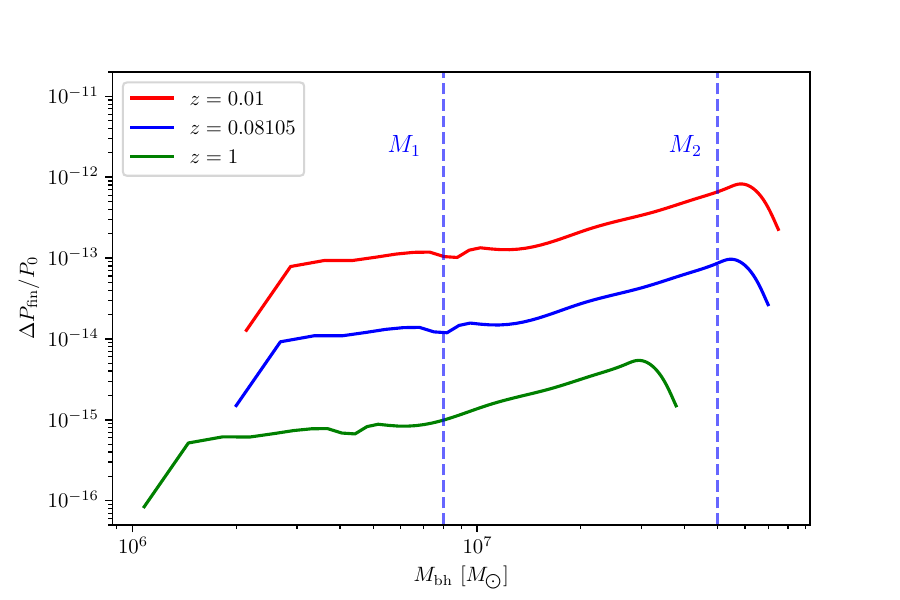}
      \caption{The relationship between $\Delta P_{\rm fin} / P_0$ and $M_{\rm bh}$ at redshifts 0.01, 0.08105 and 1.}
      \label{fig:dP_vs_m}
  \end{figure}

  For simplicity, we first consider equal-mass SMBHBs with component mass $M_{\rm bh}$. Still, for each value of $M_{\rm bh}$, we set the angular parameters to {\bf OP}. For sources at cosmological distances, $M_{\rm bh}$ enters the expression of GW in the form of redshifted mass $M_{{\rm bh}, z} \equiv (1 + z) M_{\rm bh}$. Theoretically, the influence of $M_{{\rm bh}, z}$ is twofold. Firstly, for different $M_{{\rm bh}, z}$, $f_{\rm res}$ appears at different stages of GW; secondly, the amplitude of GW is directly related to $M_{{\rm bh}, z}$. In addition, the GW amplitude is also inversely proportional to $d_L(z)$, which, at low redshifts, follows the Hubble law $d_L \propto z$. The relationships between $\Delta P_{\rm fin} / P_0$ and $M_{\rm bh}$ at redshifts $z = 0.01, 0.08105, 1$ are shown in Fig.~\ref{fig:dP_vs_m}. At the redshift of {\bf TS} ($z = 0.08105$), range $(M_1, M_2)$ marked in Fig.~\ref{fig:dP_vs_m} includes the $M_{\rm bh}$ values which would cause resonance to start at the merger stage, while if $M_{\rm bh} > M_2$, resonance occurs during the inspiral stage. A peak of $\Delta P_{\rm fin} / P_0$ can be found at $M_{\rm bh} \approx 5.5 \times 10^7 M_{\bigodot}$.
  
\begin{figure}%[H]
      \centering
      \includegraphics[width=.42\textwidth]{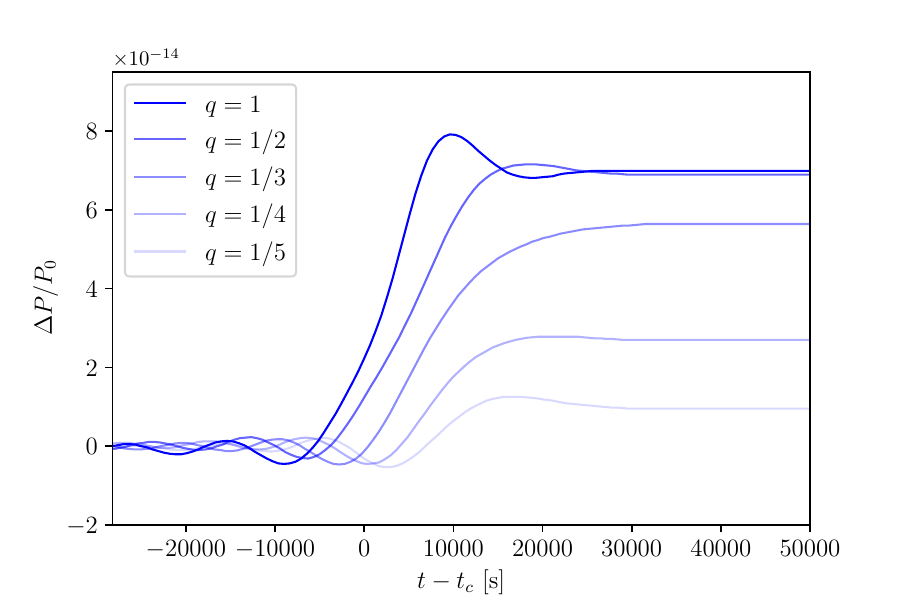}
      \caption{The optimal resonant responses of {\bf L2} for different values of $q$ with fixed $\mathcal{M}_c$.}
      \label{fig:dP_vs_q}
  \end{figure}

  The {\bf TS} is reported to be an uneven mass-ratio system, thus there is necessity to examine the role of mass ratio $q \equiv m_2 / m_1$. By varying $q$ and keeping the chirp mass $\mathcal{M}_c$ fixed, the resonant responses of {\bf L2} are plotted in Fig.~\ref{fig:dP_vs_q}. Results show that equal mass $(q = 1)$ turns out to be the most optimistic case. 

  %%%%%%%%%%%%%%%%%%%%%%%%%%%%%%%%%%%%%%%% C %%%%%%%%%%%%%%%%%%%%%%%%%%%%%%%%%%%%%%%%%%%%%%%%%%%%%%%%%%%%%%%%
  \section{The signal-to-noise ratio of distance residual}\label{sec:appendixC}
  The one-sided PSD $S_n(f)$ is defined as twice the Fourier transform of auto-correlation function $R(\tau) = \langle n(t) n(t + \tau)\rangle$, $n(t)$ being the noise at time $t$. 
  Following~\cite{PhysRevD.105.064021}, the range measurements are assumed to be unbiased, with uncorrelated Gaussian noise of variance $\sigma^2$. Thus, for discrete SLR data, by denoting the time interval between two adjacent measurements as $t_s$, $R(\tau)$ can be modeled as 
  \begin{equation}
      R(\tau) = \left\{\begin{aligned}
          \sigma^2, & \quad |\tau| < t_s / 2 \\
          0\ , & \quad  |\tau| \ge t_s / 2
          \end{aligned}\right.,
  \end{equation}
  thus 
  \begin{equation}
      S_n(f) = 2\sigma^2 t_s {\rm sinc}(\pi t_s f).
  \end{equation}
  Considering that the maximum frequency $f_{\rm max}$ of $\delta \tilde{r}(f)$ is usually much smaller than $1/t_s$, we can approximate ${\rm sinc}(\pi t_s f)$ to 1, and it follows that
  % a rectangle window function with height 1 and width $2/t_s$
  \begin{eqnarray}\label{eq:discrete_snr}
      \rho^2 &\approx& \frac{2}{\sigma^2 t_s} \int_{0}^{f_{\rm max}} |\delta \tilde{r}(f)|^2 \ df \nonumber \\
      &\approx& \frac{1}{\sigma^2 t_s} \int_0^{T_{\rm obs}} \delta r^2(t) \ dt \approx \frac{1}{\sigma^2} \sum_{i = 1}^{N_{\rm obs}} \delta r_i^2,
  \end{eqnarray}
  where $\delta r_i \equiv \delta r(t_i)$, $N_{\rm obs}$ is the total number of normal point measurements, and $T_{\rm obs} = t_s N_{\rm obs}$. Note that to derive the second line, we have used the Parseval's theorem. 
  
  For long-term data tracking, SNR is mainly contributed by the post-resonance stage. By assuming the ideal case that the only perturbation to the satellite orbit is from the incident GW of SMBHB, during the post-resonance stage, $\delta r(t)$ is the difference between two stable Keplerian orbits. We first expand $r(t)$ in terms of the Hansen coefficients to the linear order of $e$ as
  \begin{equation}
      r(t; a, e, \varepsilon) = a \left[1 - e\cos M(t; a, \varepsilon) + \mathcal{O}(e^2)\right].
  \end{equation}
  Thus the GW-induced distance residual reads 
  \begin{eqnarray}
      \delta r(t) &\approx& \frac{\partial r}{\partial a} \Delta a_{\rm fin} + \frac{\partial r}{\partial e} \Delta e_{\rm fin} + \frac{\partial r}{\partial \varepsilon} \Delta \varepsilon_{\rm fin} \nonumber \\
      &\approx& \Delta a_{\rm fin}\left[ 1 - e\left(1 + \frac{\Delta e_{\rm fin} / e}{\Delta a_{\rm fin} / a}\right)\cos M \right. \nonumber \\
      & & \quad \quad \ \  - \left. e\left(\frac{3\pi t}{P} - \frac{\Delta \varepsilon_{\rm fin}}{\Delta a_{\rm fin} / a}\right)\sin M\right],
  \end{eqnarray}
  indicating that under the long-term condition $(3\pi t / P \gg 1)$, $\delta r(t)$ would oscillate around $\Delta a_{\rm fin}$ with a linearly varying amplitude, and the rate of variation is proportional to $e$.
  The results of numerical calculation allow us to make an examination on the magnitudes of $\Delta a_{\rm fin}$, $\Delta e_{\rm fin}$ and $\Delta \varepsilon_{\rm fin}$. In the case of {\bf L2}'s optimal response, the $3\pi t / P \sin M$ term dominates over other $\mathcal{O}(e)$ terms, thus 
  \begin{equation}\label{eq:dr}
      \delta r(t) \approx \left(1 - \frac{3\pi t}{P}e\sin M \right)\Delta a_{\rm fin}.
  \end{equation}
   By inserting Eq.(\ref{eq:dr}) into Eq.(\ref{eq:discrete_snr}), the SNR of {\bf L2} can be approximated as
  \begin{equation}
      \rho^2 \approx \frac{\Delta a_{\rm fin}^2}{\sigma^2 t_s}\left(T_{\rm obs} + \frac{3\pi^2 e^2}{2P^2}T^3_{\rm obs}\right).
  \end{equation}
  % For the case shown in Fig.~\ref{fig:r_residual}, when the scale of $T_{\rm obs}$ is much smaller than $P^2 / \Delta P_{\rm fin}$, the amplitude of $\delta r$ increases linearly in time:
  % \begin{equation}
  %     \delta r(t) = k T_{\rm obs} \sin \frac{2\pi}{P_0}T_{\rm obs},
  % \end{equation}
  % where $k$ is the growth rate. 
  % Ignoring the osculating terms, the quadratic sum of residuals reads
  % \begin{equation}
  %     \sum \delta r_i^2 = \frac{k^2 t_s^2}{12}N_{\rm obs} (N_{\rm obs} + 1) (2N_{\rm obs} + 1) \approx \frac{k^2 t_s^2}{6}N_{\rm obs}^3.
  % \end{equation}
  % As a result, the relationship between ${\rm SNR}$ and $T_{\rm obs}$ can be approximated as a power-law function:   
  % \begin{equation}
  %     \rho(T_{\rm obs}) \approx \frac{k}{\sqrt{6 t_s}\sigma} T_{\rm obs}^{3/2} \propto T_{\rm obs}^{3/2}.
  % \end{equation}

% \end{appendix}

%
% BibTeX users please use
\bibliographystyle{CustomBibStyle}
\bibliography{main}
% \bibdata{main}
% \bibstyle{plain}
% \begin{thebibliography}{}
% %
% % and use \bibitem to create references.
% % Format for Journal Reference
% Author, Journal \textbf{Volume}, (year) page numbers.
% Author, \textit{Book title} (Publisher, place year) page numbers
% \end{thebibliography}

\end{document}